\documentclass[journal=jpcc, manuscript=article]{achemso}
\setkeys{acs}{articletitle = true}
\usepackage[version=3]{mhchem} 
\usepackage{txfonts}
\usepackage{graphicx}
\usepackage{cases}
\usepackage{dcolumn}
\usepackage{bm}
\usepackage{multirow}
\usepackage{float}
\usepackage{tabularx}
\usepackage{color}
\usepackage[normalem]{ulem}

\def\eV{\,\textrm{eV}}

\newcommand{\onlinecite}[1]{\hspace{-1 ex} \nocite{#1}\citenum{#1}} 

\author{Xiang-Guo Li}
\affiliation[University of Florida]
{Department of Physics and Quantum Theory Project, University of Florida, Gainesville, FL 32611}

\author{Yun-Peng Wang}
\affiliation[University of Florida]
{Department of Physics and Quantum Theory Project, University of Florida, Gainesville, FL 32611}

\author{X.-G. Zhang}
\affiliation[University of Florida]
{Department of Physics and Quantum Theory Project, University of Florida, Gainesville, FL 32611}

\author{Hai-Ping Cheng}

\email{cheng@qtp.ufl.edu}
\phone{(+1)352-392-6256}
\fax{(+1)352-392-8722}
\affiliation[University of Florida]
{Department of Physics and Quantum Theory Project, University of Florida, Gainesville, FL 32611}

\title  {Tunneling Field-Effect Junctions with WS$_2$ barrier}

\begin{document}

\begin{abstract}  
      Transition metal dichalcogenides (TMDCs), with their two-dimensional structures and sizable bandgaps, are good candidates for barrier materials in tunneling field-effect transistor (TFET) formed from atomic precision vertical stacks of graphene and insulating crystals of a few atomic layers in thickness.
We report first-principles study of the electronic properties of the Graphene/WS$_2$/Graphene sandwich structure revealing strong interface
effects on dielectric properties and predicting a high ON/OFF ratio with an appropriate WS$_2$ thickness and a suitable range of the gate voltage.
Both the band spin-orbit coupling splitting and the dielectric constant of the WS$_2$ layer depend on its thickness when in contact
with the graphene electrodes, indicating strong influence from graphene across the interfaces. The dielectric constant is significantly reduced from
the bulk WS$_2$ value. The effective barrier height varies with WS$_2$ thickness and can be tuned by a gate
voltage. These results are critical for future nanoelectronic device designs.
    
\end{abstract}

\section{\label{sec:1}Introduction}
 Performance of electronic and optoelectronic devices is fundamentally linked to the physical properties of materials that make them. The link between device performance and consequences
of quantum physics
becomes increasingly direct and apparent as devices are scaled to the nanometer range and low dimensions. A primary example is the direct link between the band structure
of semiconductors and their abilities to absorb and emit light.\cite{Gap3} The direct to indirect band gap change for few layer transition metal dichalcogenides (TMDCs), 
for example, combined with strong spin-orbit coupling and valley interactions,\cite{layTMD, Valley1,Valley2} provides tremendous potential for coupling optical, spin, and
valley electronics in these materials. 
TMDCs have been studied by researchers for decades for their electronic, optical, mechanical, and thermal properties\cite{TMD1,TMD2,TMD3,TMD4}. Recent advances in sample preparation, optical detection, transfer and manipulation of 2D materials have paved the way for their use in new field effect transistor (FET) and optoelectronic device designs. In addition to the change in band gap as a function of number of atomic layers, other properties of TDMC materials may also be dependent on
their thickness. The dielectric constant of few-layer In$_2$Se$_3$ nanoflakes has been found to be dependent on the thickness.\cite{In2Se3} 
Band splitting in TMDCs due to spin-orbit coupling, on the other hand, was found to be less sensitive to their thickness.\cite{SOWS2}
How these properties are impacted by contacting to graphene electrodes in a vertical tunnel junction structure under different layer thicknesses is an open question.

Another example of the direct impact of quantum mechanics on electronics performance is the ON/OFF ratio, 
the ratio of the electric current through a FET between its ``ON'' state and its ``OFF'' state and one of the key performance parameters of
the device. This ratio is fundamentally controlled by the band gap and the carrier mobility of the channel material. Such link is accentuated by new two-dimensional materials
represented by graphene.
Early planar graphene-based FET \cite{pFET} suffered from a small ON/OFF ratio (less than a factor of 10) due to
the absence of a band gap in graphene. Engineering a band gap in graphene using nanostructuring\cite{nano1,nano2,nano3}, chemical functionalization\cite{Chem, Chem1}, or bilayer graphene\cite{BlayG, BlayG1} adds complexity and degrades graphene's electronic quality. 
An alternative graphene-based transistor architecture is a vertical tunneling field effect transistor (TFET)\cite{vFET}. It relies on quantum tunneling\cite{QT1,QT2} through a thin insulator barrier inserted between two graphene sheets. The performance of the TFET, in particular its ON/OFF ratio, depends critically on the effective tunnel barrier height. The 2D hexagonal boron nitride ($h$-BN) has a band gap and can be easily transferred on top of graphene, but its large band gap makes
a high effective tunnel barrier that prevents the device from reaching the fully ON state, limiting the ON/OFF ratio \cite{vFET} to less than about $50$.

Several 2D TMDCs such as WS$_2$ and MoS$_2$ with band gaps around 1-2 \eV\ \cite{Gap1, Gap2,Gap3} have been experimentally demonstrated to yield ON/OFF ratios exceeding 10$^6$ when used as atomic thin tunnel barrier layers\cite{vFET1}. That a moderate band gap of 1-2 \eV\ can yield such a high ON/OFF ratio contrasts sharply with all-graphene FETs without a band gap and TFETs with a large band gap both of which have very low ON/OFF ratios. Moreover,
the TMDCs are known for their ``soft''
band gaps that can be changed easily via externally controlled parameters \cite{tgap1,tgap2,tgap3}, for example via the gate voltage and the barrier layer
thickness. Understanding how the band gaps and the barrier potential change with the
gate voltage and the layer thickness will provide us with both physical insight of these materials and guidance for how to use them in device design.

These questions are answered through a series of 
density-functional theory (DFT) \cite{DFT} calculations, in which we study 
the size (number of atomic layers of WS$_2$) and gate voltage dependences of the band splitting due to spin-orbit coupling, the dielectric constant, and the effective tunnel barrier in Graphene/WS$_2$/Graphene (Gr/WS$_2$/Gr) sanwich structures. 
To relate first-principles calculations with experimental observations, we combine the first-principles electronic structure calculations with a tunneling transport model to obtain 
the ON/OFF ratio and the current-voltage characteristics of the junction.
 
\section{\label{sec:2}Theoretical Model and Computational Details}

\begin{figure*}[t]
{\includegraphics[width=0.9\columnwidth]{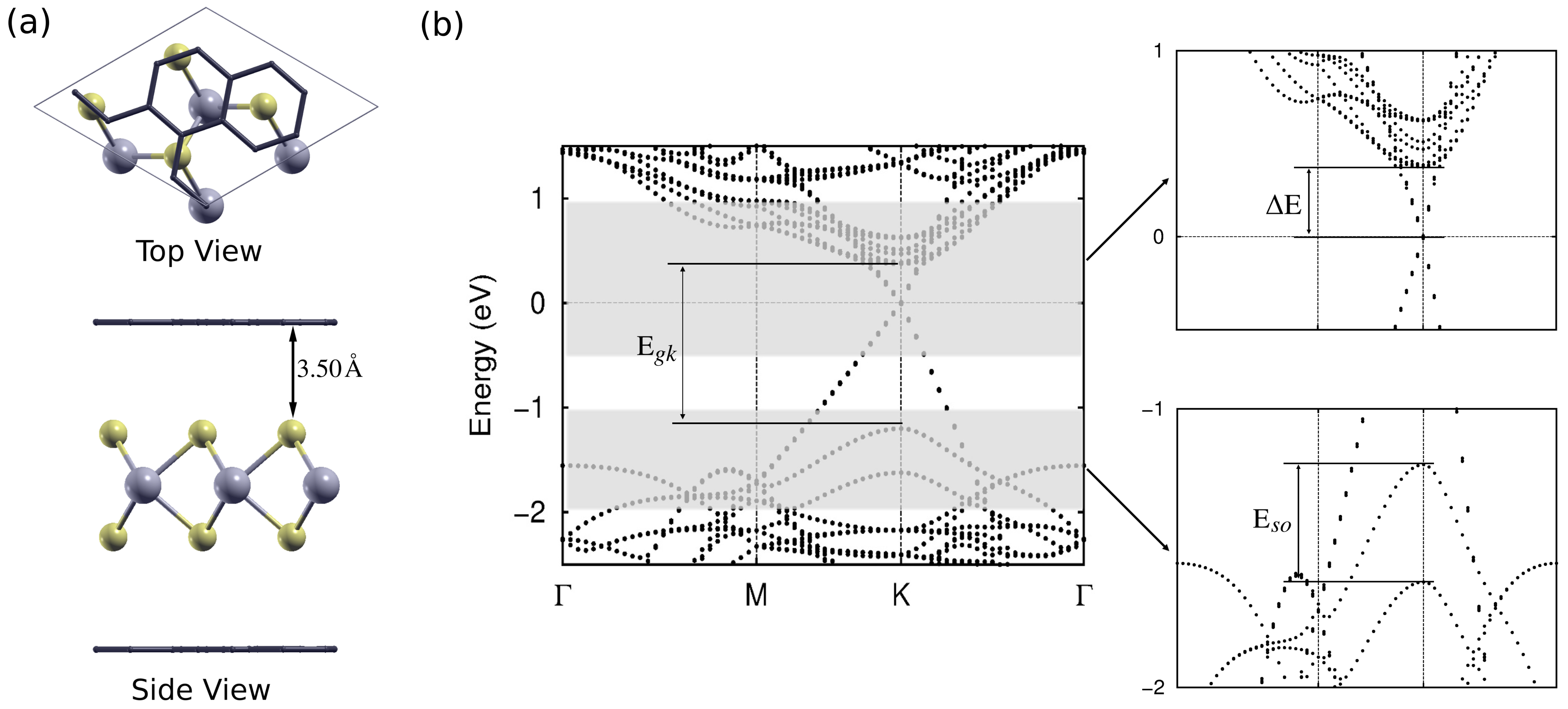}} \caption{\label{fig:str}  
Geometrical and band strucutre of Gr//WS$_2$/Gr hybrid system. \textbf{(a)} Schematic top and side views of a hybrid Gr/WS$_2$/Gr sandwich with a monolayer of WS$_2$. \textbf{(b)} Band structure of Gr/monolayer WS$_2$/Gr system along high symmetric points $\Gamma$$-$M$-$K$-\Gamma$ of 2$\times$2 WS$_2$ supercell. Two shaded energy intervals are magnified in the panels on the right. 
}
\end{figure*}

First-principles electronic structure calculations are performed based on the Perdew-Burke-Ernzerhof (PBE) exchange correlation functional to the density functional theory (DFT)\cite{DFT} as implemented in the Quantum-Espresso package\cite{QE}. We employ fully relativistic projector-augmented-wave (PAW) pseudopotentials \cite{PAW1, PAW2} including spin-orbit coupling and a plan-wave basis with the cutoff energy of 60 Ry. For structural relaxation, the optimized exchange van der Waals functional B86b of Becke (opt B86b vdW) functional\cite{VDW} is chosen to include vdW interaction. The convergence thresholds for energy and force are 10$^{-8}$ Ry and 0.0015 Ry/$\AA$, respectively. A combination of $40\times 40\times 1$ and $6 \times 6 \times 1 $ $k$-point samplings are applied for single-point energy and structure relaxation calculations, respectively. The gate field is applied using the effective screening medium (ESM) method\cite{ESM2} as implemented in Ref. \onlinecite{ESM1}.  Technically, the gate field is created by adding a suitable extra electrons or holes to Gr/WS$_2$/Gr since the areal density of free carriers is proportional to the gate voltage\cite{ESM1}. To simulate a Gr/WS$_2$/Gr sandwich structure, we consider an S-terminated surface of WS$_2$ with a triangular lattice of S atoms at the topmost and bottommost layers. A supercell, with a commensurate $\sqrt{7}\times \sqrt{7}$ lateral periodicity of graphene and $2\times2$ lateral periodicity of the WS$_2$ surface, is used  (see Fig. \ref{fig:str}a). A vacuum layer of more than 20$\,\AA$ in thickness is inserted in the direction normal to the layers between supercells. After structural relaxation, graphene layers maintain their planar and hexagonal atomic network and the average distance between the graphene layer and 
the sulfur atoms in the first WS$_2$ layer is about 3.50$\,\AA$ (see Fig. \ref{fig:str}a).

\section{\label{sec:3}Results and Discussion}
Free-standing graphene is a zero-gap semimetal whose conduction and valence bands meet at the Dirac point. Pure WS$_2$ exhibits semiconducting character with a band gap changing from direct (monolayer $\sim$2.1\eV) to indirect (bulk $\sim$1.4\eV) \cite{Gap3}. In Fig. \ref{fig:str}b we show the band structure of a Gr/monolayer WS$_2$/Gr sandwich along high symmetry $k$ directions of the WS$_2$ supercell. The Dirac point of the two graphene layers appear within the energy gap of WS$_2$, and the linear bands around the Dirac point of free-standing graphene are preserved due to the weak van der Waals (vdW) interaction between graphene and WS$_2$\cite{Gra-WS2}. Similar to a single graphene sheet, the sandwich system also has a low density of states (DOS) around the Fermi level making it easy to move the Fermi level by a gate voltage. The conduction band minimum (CBM) and the valence band maximum (VBM) of WS$_2$ at K point defines the band gap ($E_{gk}$) of WS$_2$. The energy difference between the CBM and the Fermi level is defined as the effective tunneling barrier height, $\Delta E$, an important parameter in electron tunneling for FET (see Fig. \ref{fig:str}b). The energy splitting of the valence band at K point ($E_{so}$) due to the strong spin-orbit coupling in WS$_2$\cite{SO1,SO2} is also indicated in Fig. \ref{fig:str}b.

\begin{figure*}[t]
{\includegraphics[width=0.9\columnwidth]{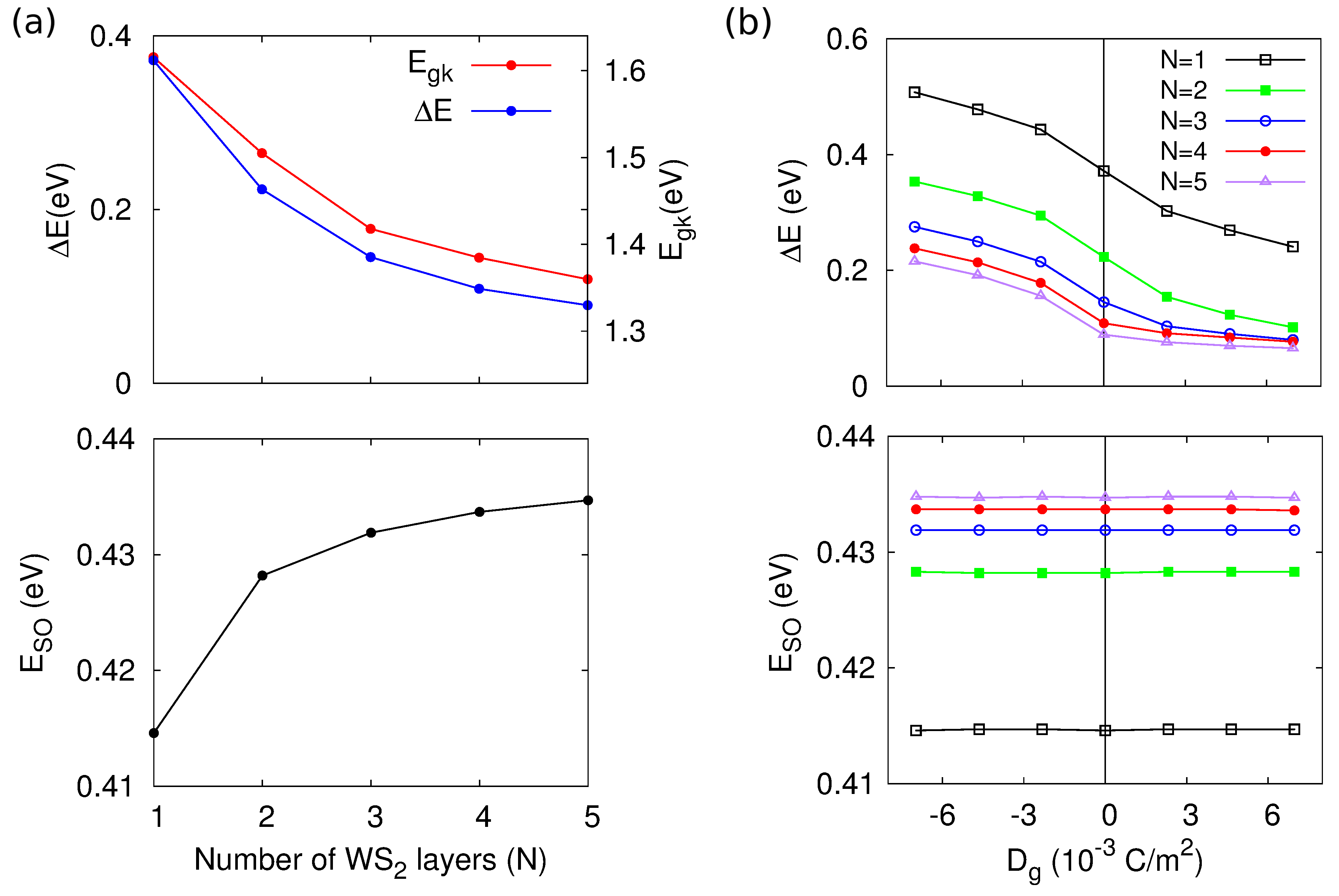}} \caption{\label{fig:data}  
The effect of gate field and WS$_2$ layer thickness on the electronic properties of the Gr/WS$_2$/Gr hybrid system. The band gap of WS$_2$ at K ($E_{gk}$), effective tunneling barrier height ($\Delta E$) and energy splittings of valence band at K due to the spin-orbit coupling in WS$_2$ as a function of \textbf{(a)} the number of WS$_2$ layers and \textbf{(b)} the gate field $D_g$ for different number of WS$_2$ layers. N donotes the number of WS$_2$ layers . }
\end{figure*}

{\bf{Thickness, Gate Field Effects and ON/OFF ratio.}}
As informed by experiment \cite{vFET1}, the ON/OFF ratio changes with the number of WS$_2$ layers in a Gr/WS$_2$/Gr junction. We first examine band structure evolution with the WS$_2$ layer thickness. Figure \ref{fig:data}a shows how $\Delta E$, $E_{gk}$ and $E_{so}$ change with the number of WS$_2$ layers. The change in the band gap $E_{gk}$ with the number of WS$_2$ layers is from the quantum confinement effect \cite{QC}. The effective tunneling barrier height $\Delta E$ trends with $E_{gk}$ (see Fig. \ref{fig:data}a), which should
affect electron tunneling. On the other hand,  the splitting of the valence band top due to the spin-orbit coupling ($E_{so}$) shows an opposite trend with the number of WS$_2$ layers.

The effect of the gate field is equivalent to doping the hybrid system. Positive gate field corresponds to electron doping and negative gate field hole doping.  The gate field modifies the barrier height ($\Delta E$) by moving the Fermi level of the system relative to the conduction band bottom of the WS$_2$ layer. Figure \ref{fig:data}b shows the evolution of $\Delta E$ as a function of the electric displacement field $D_g$ for different number of WS$_2$ layers.  Specifically, electron doping moves the Fermi level of the system closer to the CBM of the WS$_2$ layer making $\Delta$E decreased at positive gate field and vice versa.  The rate of change slightly decreases at higher field  (large $D_g$) because of the increased DOS at the Fermi energy in graphene as the Fermi energy is moved away from the Dirac point by the gate field. When the WS$_2$ layer is thin enough (N=1, 2), the change in $\Delta E$ is symmetric for positive and negative gate field. As the thickness of WS$_2$ increases, $\Delta E$ changes by a much smaller amount under a positive gate field than under a negative gate field, making the curve non-symmetric any more. This is because when the thickness of WS$_2$ increases, the Dirac point of graphene gets closer to the CBM of WS$_2$ ($\Delta E$ getting smaller, see Figure \ref{fig:data}a). Further reducing $\Delta E$ by applying a gate field can make $\Delta E$ comparable to the gaussian smearing (0.05\eV) for the energy levels (in reality, a similar effect is achieved by thermal smearing). As a result, some of the conduction states of WS$_2$ can be occupied impeding any further movement of the Fermi level. There is no such effect for negative gate field since under a negative gate field, the Fermi level moves farther away from the CBM of WS$_2$. Thus the change in $\Delta E$ under negative gate field is almost the same for different thickness of WS$_2$ (see Figure \ref{fig:data}b). We can also see that within the experimental gate field range (limited by the gate dielectric breakdown), the maximum change in $\Delta E$ is about 0.25\eV.

The gate field has little effect on $E_{so}$. This is because the two valence bands at K are both mostly $d$ states from W atoms \cite{QC} and as the gate field is applied, the shift in the potential for these states are the same. 

To relate our first-principles calculation with experiments, we employ a simple tunneling formula \cite{transport1, transport2,transport3} (see Part ``Current-Voltage Characteristics''), 
using first-principles results as input, to calculate the tunneling current as a function of the gate voltage. The obtained I-V characteristics can then be compared with experimental data. When there are four layers of WS$_2$ in the hybrid system ($N=4$), our calculated result predicts an ON/OFF ratio approaching 1000, close to experimental measurement. The calculated I-V curve compared with  experiment are included in Part ``Current-Voltage Characteristics''.

{\bf{Dielectric properties.}}
Graphene based FETs have been widely studied by combining experimental measurements and macroscopic electrostatic models \cite{vFET,model1,model2}. In these models, the dielectric barrier between the graphene layers is treated as a region with a constant dielectric permitivity equal to the bulk value. However, interfaces can have significantly different dielectric properties than the bulk \cite{interface}. 

\begin{figure}[t]
{\includegraphics[width=1.0\columnwidth]{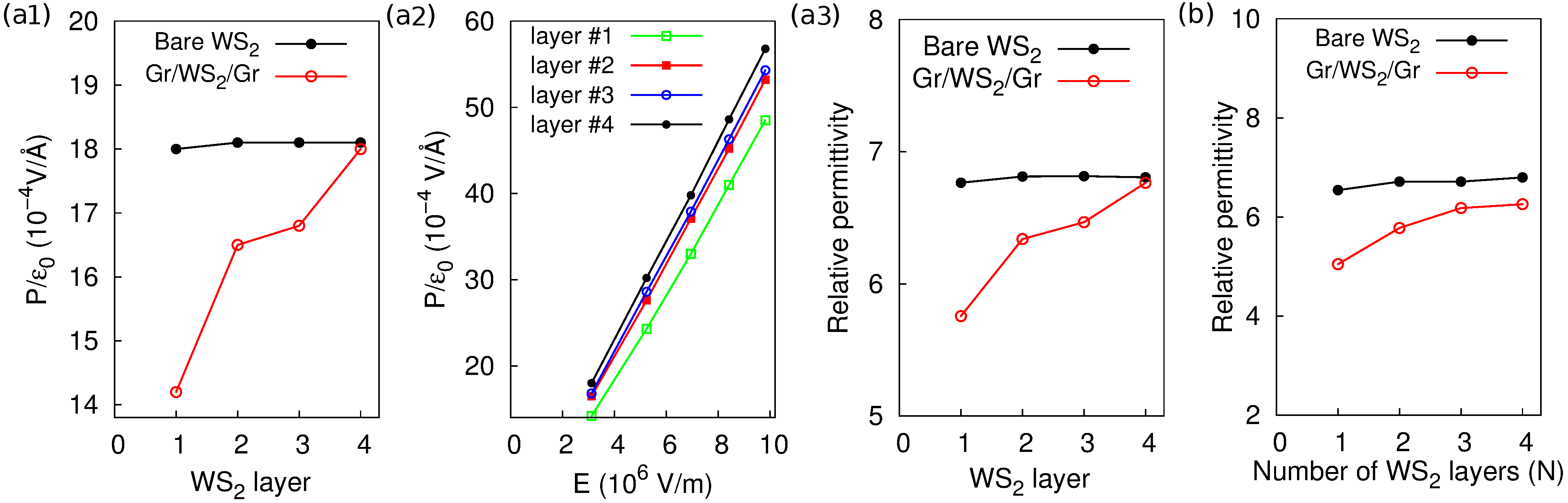}} \caption{\label{fig:die}  
Dielectric properties of the Gr/WS$_2$/Gr hybrid system compared with bare WS$_2$ system. In \textbf{(a1-a3)}, we focus on the dielectric properties of the Gr/WS$_2$/Gr hybrid system with a fixed WS$_2$ thickness (four layer WS$_2$) . The calculated local polarization in each WS$_2$ layer  \textbf{(a1)} induced by an gate field of $D_g=4.65\times 10^{-4}$\,C/m$^2$, which produces an electric field of $E=3.12\times 10^6$V$/$m within the WS$_2$ layer (red), along with data of bare WS$_2$ (without graphene electrodes, black) for comparison, and  \textbf{(a2)} as a function of the electric field ($E$) in WS$_2$.  \textbf{(a3)} The calculated local effective relative dielectric constant in each WS$_2$ layer. Layer 4 is closest to the gate. In \textbf{(b)}, we considers Gr/WS$_2$/Gr hybrid system with different WS$_2$ thickness. It shows the calculated total effective relative dielectric constant of the Gr/WS$_2$/Gr sandwich  and bare WS$_2$ (black) as a function of WS$_2$ layer thickness. }
\end{figure}

In general, the electric displacement field $D$ is related to the electric field $E$ and the polarization $P$ in the following form
\begin{eqnarray}
D=\epsilon_0E+P,\label{eq1}
\end{eqnarray}
where $\epsilon_0$ is the electric permittivity of the vacuum. $P$ can be expressed as the summation over centers of Wannier functions according to the modern theory of polarization\cite{P1,P2}, and in principle includes all dielectric properties of the medium. To calculate $P$ we follow the method in Ref.[\onlinecite{ESM1}].  The calculated polarization of the WS$_2$ layers adjacent to the graphene layers is found to be different than that of the inner WS$_2$ layers. This contrasts the case of the WS$_2$ film without the graphene
layers, in which the polarization is nearly the same across different WS$_2$ layers (see Fig. \ref{fig:die}a1), or the case of the Gr/$h$-BN/Gr system, in which the interface with graphene layers has little effect on the dielectric properties of the $h$-BN thin layers \cite{ESM1}. In the Gr/WS$_2$/Gr junction, the effect of the interface on the dielectric constant extends to several layers of WS$_2$.  

The gate field effect on the polarization of each WS$_2$ layer is shown in Fig. \ref{fig:die}a2. The linearity of all the curves allow us to extract the dielectric constant,
\begin{eqnarray}
\epsilon=1+\frac{P}{\epsilon_0E},\label{eq12}
\end{eqnarray}
where $E$ is equal to the slope of the self-consistent Kohn-Sham effective potential perpendicular to the junction. The dielectric constant in each layer of WS$_2$ extracted from this expression using the
local polarization is plotted in Fig. \ref{fig:die}a3. We can also extract a total effective dielectric constant from the total polarization. In Fig. \ref{fig:die}b we plot the total dielectric constant
as a function of the thickness of the WS$_2$ layer, and compare to that of bare WS$_2$. Since the total dielectric constant is essentially the average of the layer dielectric constant,
it shows a similar thickness dependence as that of individual layers. The result shows a significant reduction of the dielectric constant of the WS$_2$ layer due to its interface
with graphene. This contrasts sharply with the Gr/$h$-BN/Gr system, where the dielectric constant is almost layer independent \cite{ESM1}.

{\bf{Current-Voltage Characteristics.}}
The tunneling formula we employ to calculate the tunneling current explicitly includes the density of states of both the top (T) and bottom (B) graphene electrode layers,
\begin{equation}
\rho_{T(B)}(E)=\frac{2|E|}{\pi \hbar^2V_F^2}.
\end{equation}
Placing the zero-bias chemical potential of the system at $E=0$ and assuming symmetric potential shifts of $\pm eV/2$ for the two electrodes upon applying a bias voltage, the current density is given by,
\begin{eqnarray}
J(V)&=&C\int \rho_B\left(E+\frac{eV}{2}-E_{DB}\right) \rho_T\left(E-\frac{eV}{2}-E_{DT}\right)\nonumber\\
&\times& \left[f\left(E-\frac{eV}{2}\right)-f\left(E+\frac{eV}{2}\right)\right] T(E) dE,\label{eq}
\end{eqnarray}
where $C$ is a constant, 
$E_{DT(B)}$ are the energy at the Dirac point of the top (bottom) electrode measured from the chemical potential,  
\begin{equation}
f(E)=\frac{1}{e^{E/k_BT}+1}
\end{equation}
is the Fermi distribution function, and
\begin{equation}
T(E)=\begin{cases}
\exp\left[-\frac{2d_{barrier}}{\hbar}\sqrt{2m^*(\Delta E-E)}\right],  & E<\Delta E;\cr
1, & E>\Delta E.
\end{cases}
\end{equation}
is the transmission probability with an exponential form using the WKB approximations \cite{WKB}.
The effective mass $m^*$ ($\sim 0.48 \,m_0$, $m_0$ is the free electron mass) of an electron along the transport direction in WS$_2$ can be obtained from the band structure calculations. The barrier width $d_{barrier}$ is taken as the distance between the two graphene electrodes, and the effective barrier height $\Delta E$ and the energy of Dirac point $E_D$ are extracted from the first-principles calculations. The fermi velocity $V_F$ of graphene is a constant.

\begin{figure}[t]
{\includegraphics[width=0.5\columnwidth]{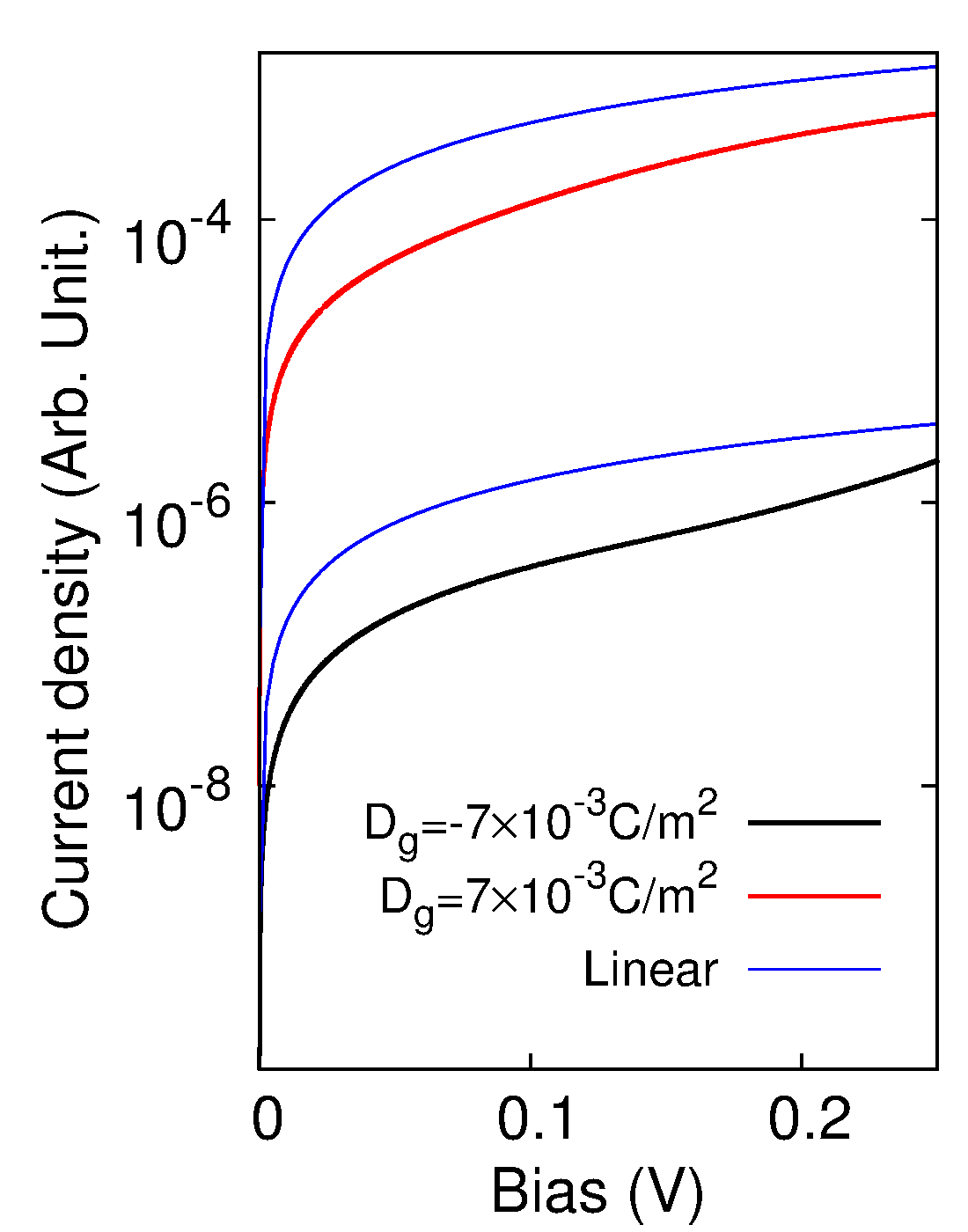}} \caption{\label{fig:Vb}  
 Current density under different gate fields $D_g$ (semi-logarithmic scale) with $N=4$. Positive (negative) $D_g$ corresponds to the ON (OFF) state. The ON state (red) curve is nearly
linear, as it closely tracks a linear function (blue curve) plotted as a guide.  $T=300$\,K .} 
\end{figure}

Equation (\ref{eq}) neglects in-plane momentum conservation, an approximation which is valid if inter-valley tunneling (i.e., incident and tranmitted wave functions belong to different
K valleys in the 2D Brillouine zone) does not make a significant contribution. The other limit in which one can neglect momentum conservation is the limit of strong interface disorder, for example
when the lattice mismatch between the layers introduces sufficient defects at the interface such that coherent momentum conserving transmission is not possible \cite{transport1}.  
Even though in both scenarios the formula takes the same form, the constant prefactor $C$ for these two cases is different. We focus on the qualitative characteristics of the I-V curve, the determination of the value of prefactor $C$ is still an open question (see supporting information).

\begin{figure}[t]
{\includegraphics[width=0.9\columnwidth]{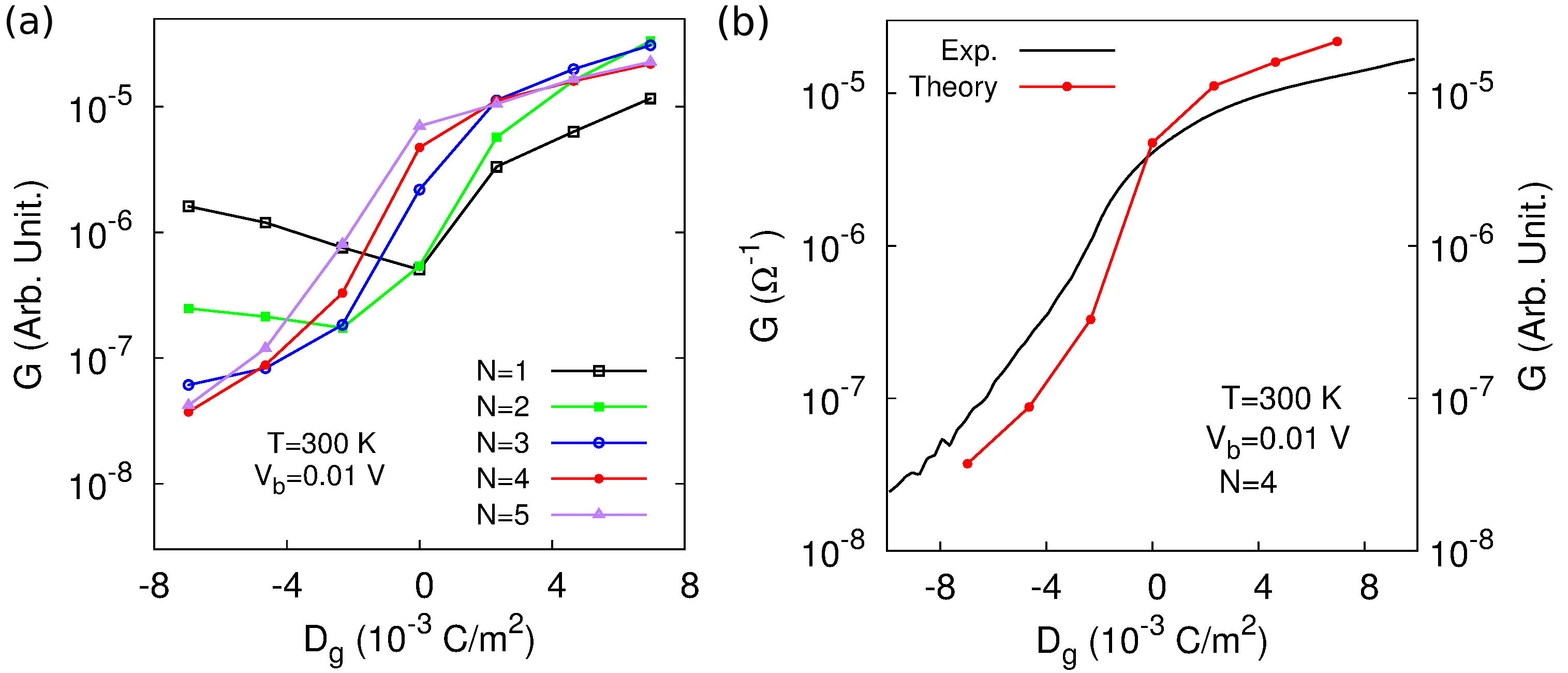}} \caption{\label{fig:transport}  
(Color online) Conductance-gate field plot at T=300\,K, bias voltage $V_b=0.01\,V$ \textbf{(a)} calculated for different number ($N$) of WS$_2$ layers, and \textbf{(b)} experimental data for $N=4$  compared to theory.  } 
\end{figure}

The above model yields the current density as a function of the bias voltage under different electric displacement fields $D_g$ (gate field) and the low bias conductance as a function of the gate field $D_g$ as plotted in Figs. {\ref{fig:Vb} and {\ref{fig:transport}}, respectively. Take as example the case of $N=4$,
for which experimental data is available. The calculated I-V curves are shown in Fig. {\ref{fig:Vb}. The device is in the ON state when the gate field is $D_g=7\times 10^{-3}C/m^2$, evident from both a large current and a nearly linear I-V curve.
A negative gate field raises the tunnel barrier and keeps the device in the OFF state as evident from greatly reduced current and a nonlinear I-V curve.  
The low bias conductance as a function of gate field $D_g$ and number of layers of WS$_2$ in Fig. {\ref{fig:transport}a} shows different trends for positive and negative gate fields. 
At negative gate fields, the barrier is high so the dominant variable in the conductance is the WS$_2$ thickness, which causes the conductance to decrease with the increasing thickness.
At positive gate fields, the barrier height is greatly reduced and increasing WS$_2$ thickness causes further reduction of the barrier height when WS$_2$ layer is thin ($N=1$ and $N=2$) leading to an increase in the conductance. This trend is reversed for $N>2$, because at these thicknesses the Fermi level is so close to the CBM of WS$_2$ (comparable to the gaussian smearing), increasing WS$_2$ thickness can not further reduce the barrier height (the change of $\Delta E$ is asymmetric in Figure \ref{fig:data}b) under positive gate fields.
For very thin WS$_2$ thicknesses ($N=1$ and $N=2$), a third competing factor, the DOS of the two graphene electrodes, comes into play. As the gate field is increased from zero in either positive or negative directions, the Fermi energy DOS increases linearly from zero. Therefore the low bias conductance also increases if all other factors remain
unchanged. This is indeed the case for $N=1$, which shows a turning point at $D_g=0$ in Fig. {\ref{fig:transport}a}, and somewhat true for $N=2$, with a turning point at a slightly negative $D_g$.  

To compare with experimental data, we plot both simulation and experimental curves for $N=4$ in Fig. \ref{fig:transport}b. The two curves agree very well.

\section{\label{sec:4}CONCLUSION}

First-principles calculations of
the electronic structures of Gr/WS$_2$/Gr field-effect transistors show that the effective barrier height of WS$_2$  in contact with graphene decreases with increasing number of WS$_2$ layers, and can be tuned by a gate field. A transport model built on top of the first-principles calculations show that the ON/OFF ratio is size-dependent and reaches a significant high value with an appropriate number of WS$_2$ layers and a suitable range of gate voltages.  The spin-orbit effect in WS$_2$ has also been investigated and it has a slight layer dependence, but is independent of the gate field. Comparison of the dielectric constant of bare WS$_2$ and Gr/WS$_2$/Gr sandwiches shows that interfacing with graphene can significantly reduce the dielectric constant of WS$_2$ and the effect gradually decreases as the number of WS$_2$ layers increases.
The effects of the barrier layer thickness and the gate field on the transport properties of Gr/WS$_2$/Gr FET are clearly connected to the band structure change
as shown from first-principles calculations, providing important information for future nanoelectronic device design.

~

\begin{acknowledgement}

  This work was supported by the US Department of Energy (DOE), Office of Basic Energy Sciences (BES), under Contract No.~DE-FG02-02ER45995. 
The computation was done using the utilities of the National Energy Research Scientific Computing Center (NERSC).\\

\end{acknowledgement}

{\bf{Supporting Information Available:}}
This material is available free of 
charge 
via the Internet at http://pubs.acs.org
.


\providecommand{\latin}[1]{#1}
\providecommand*\mcitethebibliography{\thebibliography}
\csname @ifundefined\endcsname{endmcitethebibliography}
  {\let\endmcitethebibliography\endthebibliography}{}

\end {document}